\documentclass[12pt]{iopart}
\expandafter\let\csname equation*\endcsname\relax
 \expandafter\let\csname endequation*\endcsname\relax
\usepackage{dcolumn}
\usepackage{bm}
\usepackage{amsmath}
\usepackage{amsfonts}   
\usepackage{graphicx}
\usepackage{subfigure}
\usepackage{hyperref}
\usepackage{color}
\usepackage{float}
\begin{document}
\title{Sublinear but Never Superlinear Preferential Attachment by Local
  Network Growth}
\author{Alan Gabel and S. Redner}
\address{Center for Polymer Studies and Department of Physics, Boston University, Boston,
MA 02215, USA}

\begin{abstract}

  We investigate a class of network growth rules that are based on a
  redirection algorithm wherein new nodes are added to a network by linking
  to a randomly chosen target node with some probability $1-r$ or linking to
  the parent node of the target node with probability $r$.  For fixed
  $0<r<1$, the redirection algorithm is equivalent to linear preferential
  attachment.  We show that when $r$ is a decaying function of the degree of
  the parent of the initial target, the redirection algorithm produces
  sublinear preferential attachment network growth.  We also argue that no
  local redirection algorithm can produce superlinear preferential
  attachment.

\end{abstract}

\section{Introduction}
A popular and highly successful mechanism to account for the growth of
complex networks is preferential
attachment~\cite{yule,simon,merton,price,BA99,N03}.  Here new nodes are added
sequentially to a network and each links to existing nodes according to an
attachment rate $A_k$ that is an increasing function of the degree $k$ of the
`target' node to which linking occurs.  Many real world networks appear to
evolve according to this simple
dynamics~\cite{AB02,DM03,NBW06,BBV08,N10,EL03}.

Preferential attachment networks naturally divide into three classes:
sublinear, linear, and superlinear, in which the attachment rate grows with
the degree of the target node as $A_k\sim k^\gamma$, with $0<\gamma<1$,
$\gamma=1$, and $\gamma>1$, respectively~\cite{KR01,DMS00,BRST01}.  Each class of
networks has qualitatively different properties.  Linear preferential
attachment is the most well studied class.  This growth rule leads to
scale-free networks, but with a \emph{fragile} power-law degree distribution.
Here, the term fragile denotes that the exponent of the degree distribution
depends sensitively on microscopic details of the growth mechanism.  In
contrast, sublinear preferential attachment leads to networks with a
universal stretched-exponential degree distribution.  Superlinear
preferential attachment networks are singular in character, as they contain
one highly-connected ``hub'' node whose degree is of the order of the total
number of nodes in the network~\cite{GBW04,OS05,KK08,KRB10}.

A basic feature of the preferential attachment growth rule is that each new
node must `know' the degree distribution of the entire network, as this
global information is exploited to determine the identity of the target node.
However, since real-world networks are typically large, it is unreasonable to
expect that any new node has such global knowledge.  A possible resolution is
to use a redirection rule in which a new node only knows about some local
portion of the network and attaches to a node in this region.  As we will
discuss, redirection generates linear preferential attachment growth by
suitably-defined local growth rules~\cite{KR01, klein,V03}.  An attractive
feature of redirection is that this algorithm leads to extremely efficient
simulations of network growth~\cite{KR02,MM05,CKBBL12}.

In this work, we introduce an extension of the redirection algorithm that
produces networks that grow according to \emph{sublinear} preferential
attachment.  In classic redirection, each new node attaches to either: (i) a
randomly chosen target node with probability $1-r$, where $r$ is a fixed
number strictly between 0 and 1, or (ii) the parent of the target, with
probability $r$.  We will show that a related redirection algorithm leads to
sublinear preferential attachment growth when $r$ is a suitably-chosen
decreasing function of the degree of the parent node.  Thus sublinear
preferential attachment can also be achieved from a local growth rule.
Furthermore, we will demonstrate that no local redirection algorithm can
produce superlinear preferential attachment.  Our claim may help explain why
linear and sublinear preferential attachment networks are found ubiquitously
in empirical studies~\cite{AB02, DM03, NJB03, NFB02, ML07, RLH10}, while
evidence for superlinear attachment networks are relatively
scarce~\cite{GBW04, GS12}: most real-world networks truly \emph{do} grow
according to local rules and thus cannot be governed by superlinear
preferential attachment.

\section{Preferential Attachment}
We begin with a summary of well-known basic aspects of preferential
attachment and then describe how to achieve this type of network growth by
redirection.  Let $N$ be the total number of nodes in a growing network and
$N_k$ be the number of nodes of degree $k$.  For simplicity we consider
directed tree-like networks, in which each new node has a single outgoing
link.  Thus a node of degree $k$ will have a single parent (the node it first
attached to upon entering the network) and $k-1$ children.  We assume the
network begins with a single node, which is thus its own parent.  More
general initial conditions give the same results when number of network nodes
is sufficiently large.  The network grows by introducing new nodes into the
network sequentially.  Each new node attaches to a pre-existing node of
degree $k$ with attachment rate $A_k$.

The evolution of $N_k$ with the addition of each node is thus governed
by~\cite{KR01,KRB10,AB02}
\begin{equation}
\frac{dN_k}{dN}=\frac{A_{k-1}N_{k-1}-A_kN_k}{A}+\delta_{k,1}\,,
\label{masterPA}
\end{equation}
where $A=\sum_j A_jN_j$ is the total attachment rate.  The first term
$A_{k-1}N_{k-1}/A$ gives the probability that the new node attaches to a
pre-existing node of degree $(k-1)$; this connection converts the node to
have degree $k$, thereby increasing $N_k$ by $1$.  Similarly, the term
$A_kN_k/A$ corresponds to the probability that the new node connects to a
node of degree $k$, thereby decreasing $N_k$ by $1$.  Finally, the term
$\delta_{k,1}$ arises because every new node has degree 1 and so increases
$N_1$ by 1.

In sublinear and linear preferential attachment, it can be shown~\cite{KR01}
that the total attachment rate scales as $A=\mu N$, where $\mu$ is a
constant.  In contrast, for superlinear attachment, the total rate scales as
$A\sim N^\gamma$ because $A=\sum_j A_jN_j$ is dominated by the term in the
sum that is associated with the hub, whose degree is of order $N$.  The
formal solution for $N_k$ can be readily obtained by writing $N_k=Nn_k$ and
using $A=\mu N$ in \eqref{masterPA} to give~\cite{KR01}
\begin{equation}
{n_k}=\frac{\mu}{A_k}\prod_{j=1}^{k}\left(1+\frac{\mu}{A_j}\right)^{-1}.
\label{solvePA}
\end{equation}
For sublinear preferential attachment, the asymptotic behavior may be found
by writing the above product as the exponential of the sum of a logarithm,
converting the sum to an integral, expanding the logarithm in inverse powers
of $k^\gamma$, and then performing the integrals.  These manipulations lead
to a degree distribution that asymptotically has the stretched exponential
form
\begin{equation}
\label{nk-soln}
n_k\sim k^{-\gamma}\exp\left[-\frac{\mu}{1-\gamma}\,\, k^{1-\gamma}\right]\,.  
\end{equation}
For linear preferential attachment, and more generally for shifted linear
attachment, where $A_k=k+\lambda$, the asymptotic degree distribution has the
\emph{non-universal\/} power-law form $n_k\sim k^{-3-\lambda}$n~\cite{KR01}.
Here $\lambda$ must satisfy the constraint $\lambda>-1$; otherwise, network
evolution is pathological because it is not possible to attach to nodes of
degree 1.

\section{Generalized Redirection Algorithm}

We now discuss how a suitably-defined redirection algorithm leads to a
complex network whose growth is governed by \emph{sublinear} preferential
attachment.  We assume that the network starts in a configuration in which
each node has one parent.  Let us first review the classic redirection
algorithm that leads to shifted linear preferential attachment~\cite{KR01}.
The steps to add a new node to the network are as follows:
\begin{enumerate}
\item Randomly select an existing node as the target.
\item A new node attaches to the target with probability $1-r$, with $0<r<1$.
\item With probability $r$, the new node attaches to the parent of target.
\end{enumerate}
This growth rule is local because each new node only needs to know about a
single randomly-chosen node and its immediate environment, rather than the
entire network structure.  This redirection algorithm can be
straightforwardly extended to allow the new node to make connections to
multiple nodes in the network~\cite{caveat}.  The surprising aspect of this
innocuous redirection mechanism is that it leads precisely to linear
preferential attachment for the case $r=\frac{1}{2}$~\cite{KR01}, a growth
rule that ostensibly requires knowing the degrees of all nodes in the
network.

\begin{figure}[ht]
\begin{center}
\includegraphics[width=0.65\textwidth]{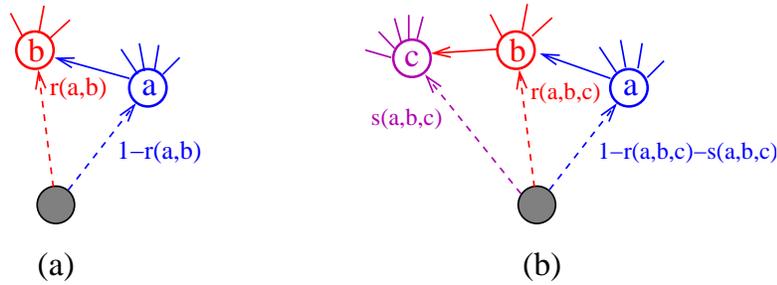} 
\caption{(a) Illustration of generalized redirection.  Attachment of a new
  node (shaded) to the ancestor (of degree $b$) of a random target (of degree
  $a$) occurs with probability $r(a,b)$, while attachment to the target
  occurs with probability $1-r(a,b)$.  (b) Redirection that extends to the
  grandparent node (of degree $c$). }
  \label{r}
\end{center}
\end{figure}

We now generalize the redirection algorithm to allow $r=r(a,b)$ to be a
function of the degrees of the target and ancestor nodes, $a$ and $b$,
respectively (Fig.~\ref{r}).  To show how sublinear preferential attachment
can be achieved from this still-local information, let us define $f_k$ as the
total probability that an incoming link is redirected \emph{from} a
randomly-selected target node of degree $k$ to the parent of the target.
Similarly, we define $t_k$ as the total probability that an incoming link is
redirected \emph{to} a parent node of degree $k$ after the incoming node
initially selected one of the child nodes of this parent.  Formally, these
probabilities are defined in terms of the redirection probabilities by
\begin{equation}
f_k=\sum_{b\geq 1}\frac{r(k,b)N(k,b)}{N_k}\,,\qquad \qquad 
t_k=\sum_{a\geq 1}\frac{r(a,k)N(a,k)}{(k-1)N_k}\,,
\label{prob}
\end{equation}
where $N_k=\sum_{b\geq 1} N(k,b)$ and $N(a,b)$ is the correlation function
that specifies the number of nodes of degree $a$ that have a parent of degree
$b$.  Thus $f_k$ is the mean redirection probability averaged over all $N_k$
possible target nodes of degree $k$.  Likewise, since each node of degree $k$
has $k-1$ children, there are $(k-1)N_k$ possible target nodes whose
redirection probabilities are averaged to give $t_k$.

In terms of these probabilities $f_k$ and $t_k$, the master equation that
governs the evolution of $N_k$ is
\begin{equation}
\label{masterRA}
\frac{dN_k}{dN}=\frac{(1\!-\!f_{k\!-\!1})N_{k-1}-(1\!-\!f_k)N_k}{N} 
+\frac{(k\!-\!2)t_{k-1}N_{k\!-\!1}-(k-1)t_{k}N_{k}}{N} +\delta_{k,1}.
\end{equation}
The first ratio corresponds to instances of the growth process for which the
incoming node actually attaches to the initial target node.  For example, the
term $(1-f_k)N_k/N$ gives the probability that one of the $N_k$ target nodes
of degree $k$ is randomly selected and that the link from the new node is
\emph{not} redirected away from this target.  Similarly, the second ratio
corresponds to instances in which the link to the target node \emph{is}
redirected to the parent.  For example, the term $(k-1)t_kN_k/N$ gives the
probability that one of the $(k-1)N_k$ children of nodes of degree $k$ is
chosen as the target and that the new node \emph{is} redirected.  Lastly, the
term $\delta_{k,1}$ accounts for the newly-added node of degree 1.

By simple rearrangement, we can express the master equation \eqref{masterRA}
in the generic form of~\eqref{masterPA}, with attachment rate given by
\begin{equation}
\label{Ak}
\frac{A_k}{A}=\frac{(k-1)t_k+1-f_k}{N}\,.
\end{equation}
As a simple check of this expression, note that when the redirection
probability is constant, we have $f_k=t_k=r$ and the expected linear
dependence $A_k\sim k$ is recovered.

\section{Sublinear Preferential Attachment}
The asymptotic behavior of the attachment rate in \eqref{Ak} is $A_k\sim
k\,t_k$.  Thus a redirection probability $r(a,b)$ for which $t_k$ is a
decreasing function of $k$ will asymptotically correspond to sublinear
preferential attachment.  Let us therefore choose
\mbox{$r(a,b)=b^{\gamma-1}$}, with $0<\gamma<1$.  Because $r$ depends only on
the degree of the parent node, Eq.~\eqref{prob} reduces to
$t_k=k^{\gamma-1}$.  Using this form of $t_k$ in Eq.~\eqref{Ak} yields
\begin{equation}
\label{SubLinear}
\frac{A_k}{A}=\frac{k^{\gamma}-k^{\gamma-1}+1-f_k}{N}
\end{equation}
whose leading behavior is indeed sublinear preferential attachment: $A_k\sim
k^{\gamma}$.  Because $f_k$ is a bounded probability, it represents a
subdominant contribution to $A_k$.

\begin{figure}[ht]
\begin{center}
\begin{tabular}{c}
\includegraphics[width=0.475\textwidth]{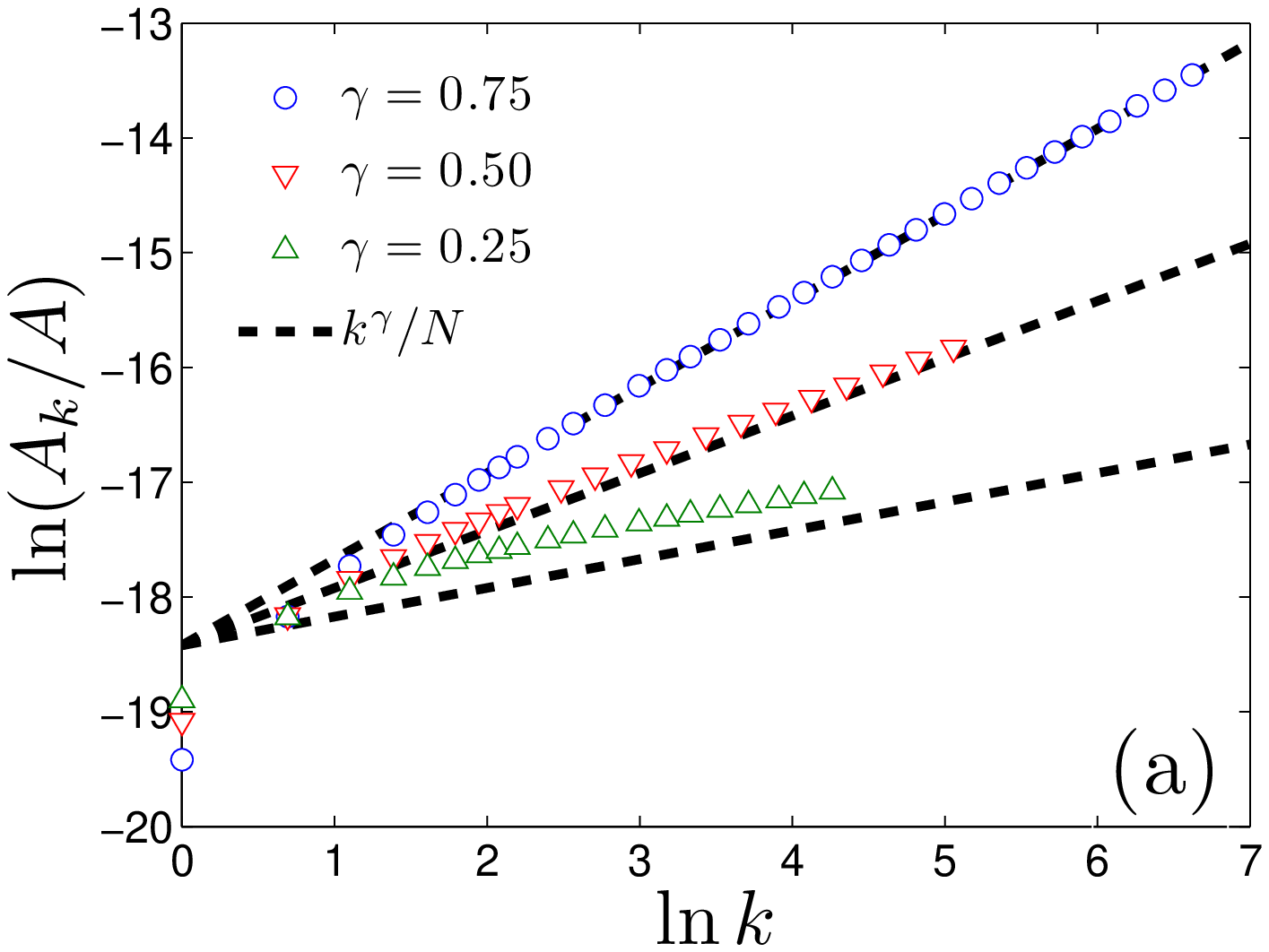} 
\includegraphics[width=0.475\textwidth]{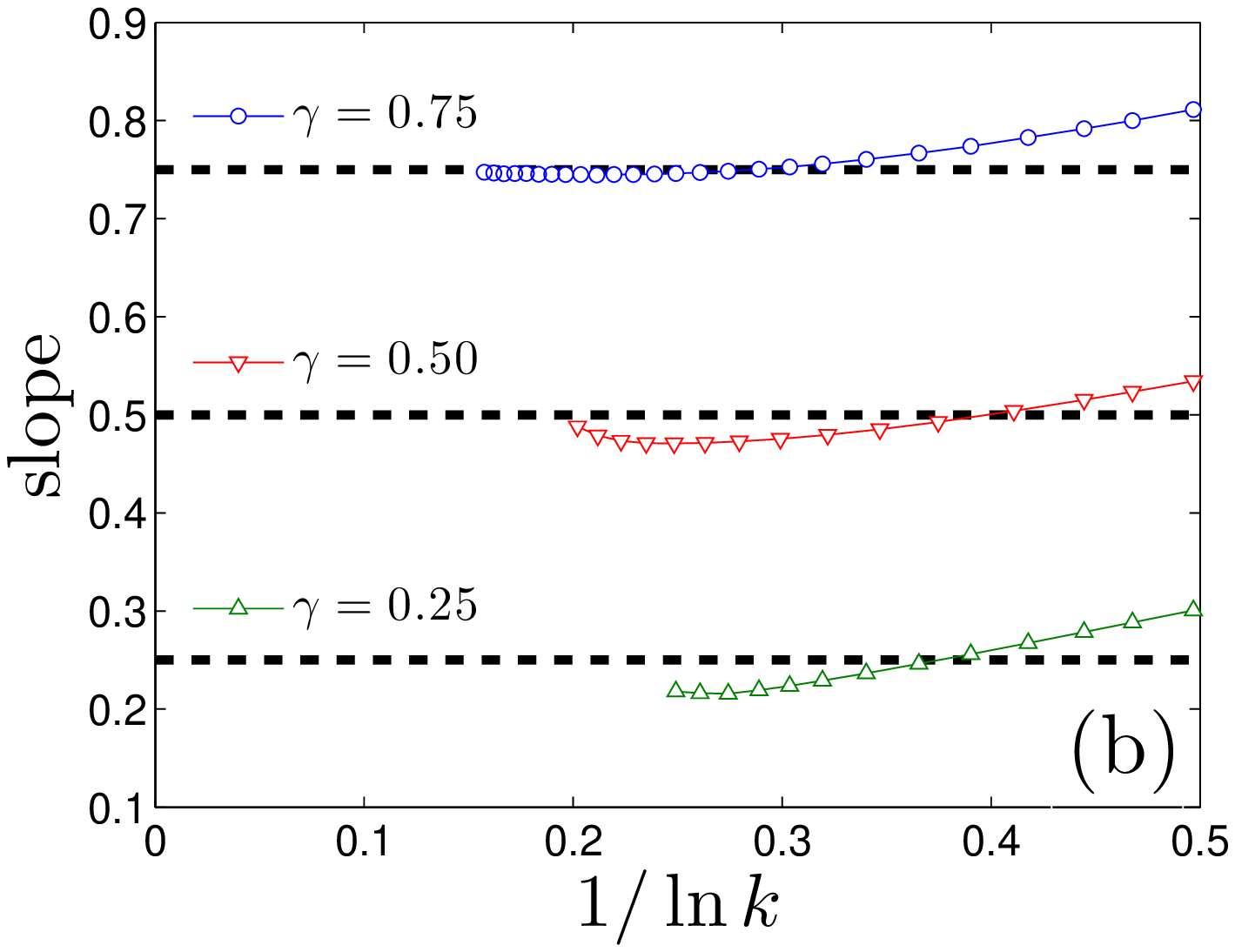}
\end{tabular}
\caption{(a) Simulated attachment probabilities $A_k/A$ versus $k$ from
  generalized nearest-neighbor redirection.  (b) Local slopes of the
  successive data points in (a) versus $1/\ln k$ for $\gamma=0.75$ ($\circ$),
  $\gamma=0.5$ ($\nabla$), and $\gamma=0.25$ ($\Delta$).  The dashed curve
  represents $A_k/A=k^\gamma/N$.}
  \label{FigAk}
\end{center}
\end{figure}

To test the prediction that $A_k\sim k^{\gamma}$ with $\gamma<1$, we
simulated 100 network realizations that are grown to $N=10^8$ nodes by
generalized redirection.  Once a network reaches $10^8$ nodes, we measure the
probability that attachment to a node of degree $k$ actually occurs by
systematically making `test' attachments to each node of the network
according to generalized redirection.  The term test attachment means that
the network is returned to its original state after each such event.  We
count all test events that ultimately lead to attachment to a node of degree
$k$.  Dividing this number of events by the total number of nodes $N$ gives
the attachment probability to nodes of degree $k$, $A_kN_k/A$.

\begin{figure}[htb]
\begin{center}
\includegraphics[width=0.475\textwidth]{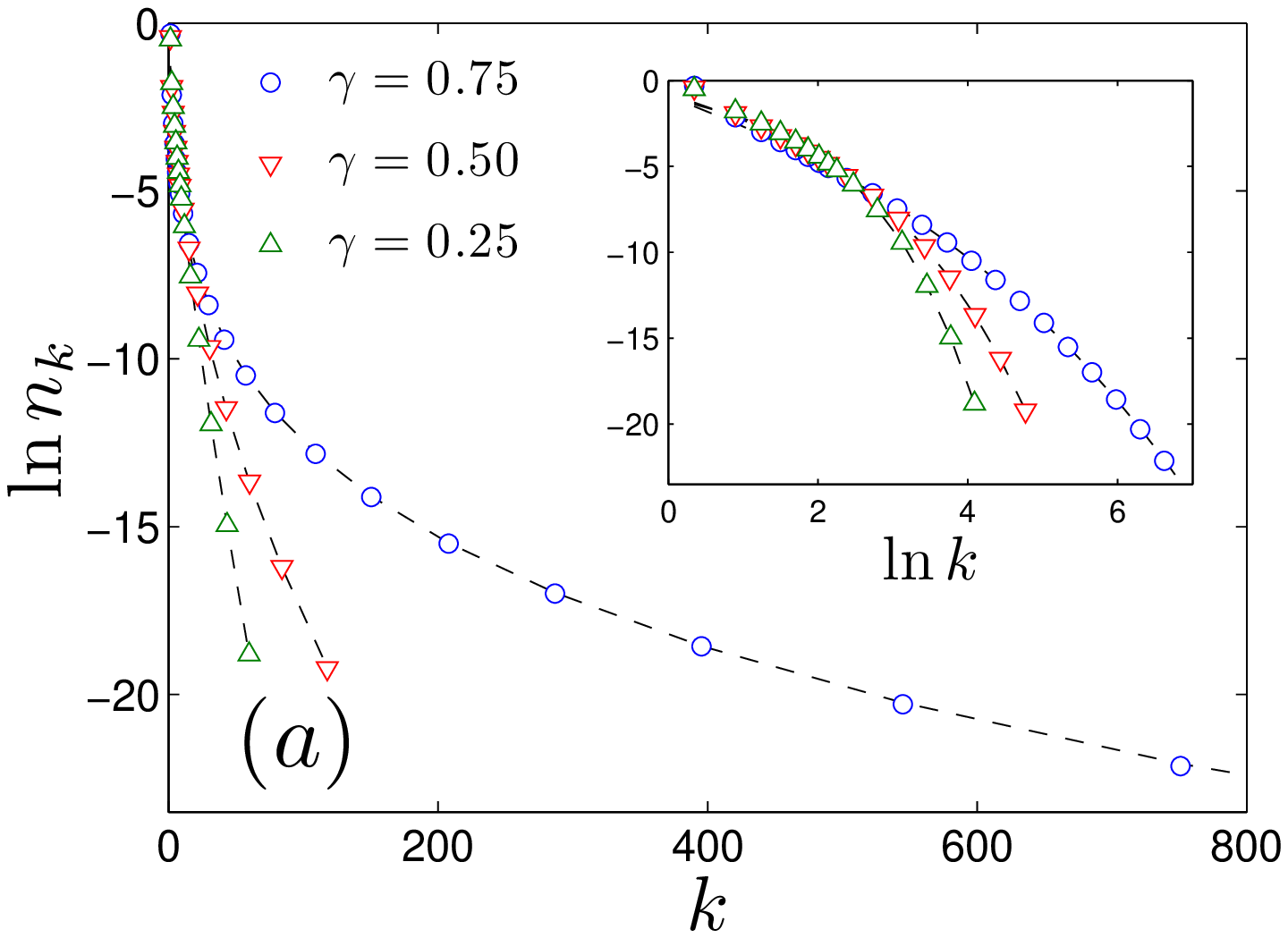}
\includegraphics[width=0.475\textwidth]{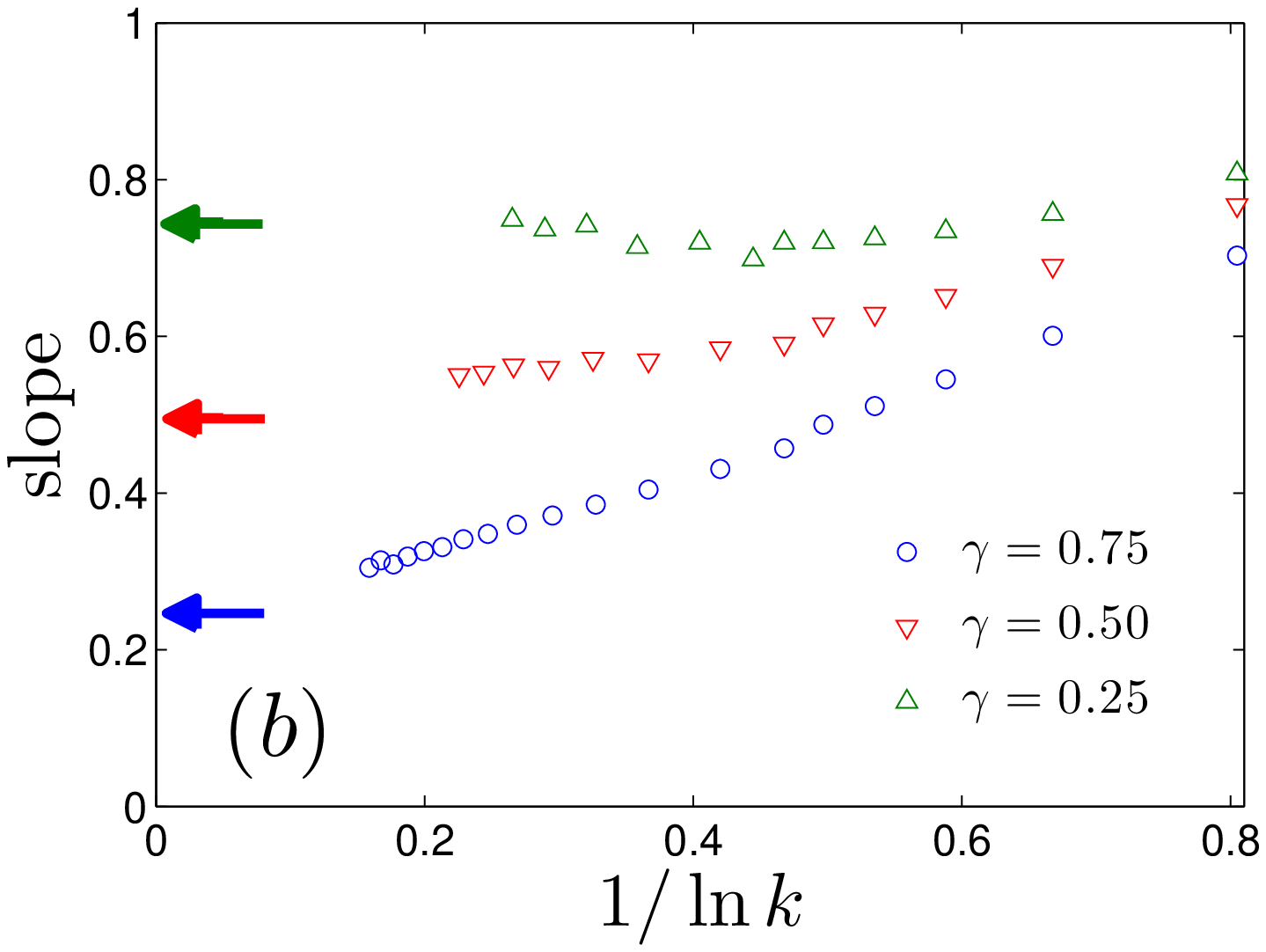}
\caption{(a) Degree distribution $n_k$ versus $k$ for generalized redirection
  with $r(a,b)=b^{\gamma-1}$ for $\gamma=0.75$ ($\circ$), $\gamma=0.5$
  ($\bigtriangledown$), and $\gamma=0.25$ ($\bigtriangleup$).  The data are
  accumulated in equal-size bins on a logarithmic scale.  Dashed curves are
  least-squares fits based on Eq.~\eqref{fit} for $k\ge 4$ with parameters
  $(C,\tilde{\mu})=(0.45,0.99)$ for $\gamma=0.75$, $(0.59,0.90)$ for
  $\gamma=0.5$, and $(0.59,0.75)$ for $\gamma=0.25$.  Inset: the same data on
  a double logarithmic scale. (b) The local slopes from a plot of
  $\ln[\ln(k^\gamma n_k)]$ versus $\ln k$; the arrows show the expected
  asymptotic result from Eq.~\eqref{nk-soln}, $\ln{k^\gamma n_k} \sim
  -k^{1-\gamma}$.}
\label{nk}
\end{center}
\end{figure}

Our simulations indeed show that the $A_k$ grows sublinearly with
$k$~(Fig.~\ref{FigAk}).  The agreement is best for $\gamma$ less than, but
close to 1, where the degree distribution is fairly broad.  As $\gamma$
decreases, the asymptotic behavior is contaminated by the appearance of
progressively more slowly-decaying sub-asymptotic corrections terms in
Eq.~\eqref{Ak}.  We also compare the simulated degree distribution to the
analytic result given in Eq.~\eqref{solvePA} with $A_k=k^\gamma$.  To make
this comparison, we fit the simulated distribution to
\begin{equation}
  {n_k}=C\,\,\frac{\tilde{\mu}}{k^\gamma}\,
  \prod_{j=1}^{k}\left(1+\frac{\tilde{\mu}}{j^\gamma}\right)^{-1},
\label{fit}
\end{equation}
where $C$ and $\tilde{\mu}$ are fitting parameters.  We introduce these
parameters because our redirection algorithm gives the attachment rate
$A_k\sim k^\gamma$ only asymptotically.
Thus the degree distribution that arises from our generalized redirection
algorithm should match the theoretical prediction \eqref{solvePA} only as
$k\to\infty$~(Fig.~\ref{nk}).  While the best-fit value of $\tilde{\mu}$ does
not obey the bound $\mu>1$ for sublinear preferential attachment~\cite{KR01},
the fitting parameters do not affect the nature of the dependence of $n_k$ on
$k$ and $\gamma$.

\section{Unattainability of Superlinear Preferential Attachment}
From Eq.~\eqref{Ak}, we see that $A_k$ will be superlinear in $k$ only if
$t_k$ can grow as a power law in $k$.  Since $t_k$ is a probability that must
be less than one, any nearest-neighbor redirection algorithm cannot produce
superlinear preferential attachment.  In redirection, an incoming node can
attach to an arbitrary node $\mathbf{x}$ either directly or by attaching to
one of the $k-1$ children of $\mathbf{x}$ and then redirecting to
$\mathbf{x}$.  Thus there are a maximum $k$ ways for the new node to attach
to node $\mathbf{x}$.  Redirection cannot provide any additional
$k$-dependent amplification beyond the local environment size because the
factor $t_k$ in \eqref{Ak} is bounded as $k\to\infty$.  Therefore the
attachment rate to $\mathbf{x}$ cannot grow faster than linearly in its
degree by nearest-neighbor redirection.

The inability of nearest-neighbor redirection to produce superlinear
preferential attachment suggests attaching to more distant ancestors of a
node.  It is natural to consider grandparent redirection~\cite{BK09} in which
the incoming node attaches to a randomly-selected target node with
probability $1-r(a,b,c)-s(a,b,c)$, to the parent of the target with
probability $r(a,b,c)$, and to the grandparent of the target with probability
$s(a,b,c)$, where $a$, $b$, and $c$ are the degrees of the target, parent,
and grandparent nodes, respectively (Fig.~\ref{r}).

Following the same steps as in our parent redirection algorithm, $f_k$ and
$t_k$ are again the respective probabilities that an incoming link is: (a)
redirected \emph{from} a randomly selected target of degree $k$, and (b)
\emph{to} a parent node of degree $k$ from an initially selected child node.
We also introduce $u_k$ as the probability a link is redirected to the
\emph{grandparent\/} of degree $k$ from an initially selected grandchild of
this node.  These probabilities are formally defined as
\begin{align}
\begin{split}
\label{GAprob}
f_k&=\sum_{b\geq 1}\sum_{c\geq 1}\frac{\left[r(k,b,c)+s(k,b,c)\right]N(k,b,c)}{N_k}~, \\
t_k&=\sum_{a\geq 1}\sum_{c\geq 1}\frac{r(a,k,c)N(a,k,c)}{(k-1)N_k}~, \\
u_k&=\sum_{a\geq 1}\sum_{b\geq 1}\frac{s(a,b,k)N(a,b,k)}{g_kN_k}~,
\end{split}
\end{align}
where the correlation function $N(a,b,c)$ is the number of nodes of degree
$a$ with a parent of degree $b$ and a grandparent of degree $c$.  As in
nearest-neighbor redirection, $f_k$ is the total redirection probability
averaged over all $N_k$ nodes of degree $k$ and $t_k$ is the parent
redirection probability averaged over all $(k-1)N_k$ children whose parents
have degree $k$.  The function $g_k$ is defined as the mean number of
grandchildren of a node of degree $k$, and therefore, $u_k$ is the
redirection probability to a grandparent of degree $k$, averaged over all
$g_kN_k$ grandchildren of this degree-$k$ grandparent node.

It is convenient to define $g_k$ in terms of $c_k$, the mean degree of a
child of a degree-$k$ node.  A node of degree $k$ has $k-1$ children, and
each child has $c_k-1$ children, on average.  Therefore, a node of degree $k$
has, on average, $g_k=(k-1)(c_k-1)$ grandchildren.  Alternatively, we can
start with the formal definitions of $g_k$ and $c_k$:
\begin{equation}
\label{ck}
g_k=\sum_{a\ge1}\sum_{b\ge 1} N(a,b,k), \qquad\qquad c_k=\sum_{a\ge 1}aN(a,k).
\end{equation}
Now using the identities $\sum_aN(a,b,c)=(b-1)N(b,c)$ and
$\sum_aN(a,b)=(b-1)N_b$, we rearrange \eqref{ck} to also give the relation
$g_k=(k-1)(c_k-1)$.

We may now write the master equation that describes the evolution of $N_k$
in grandparent redirection
\begin{align}
\begin{split}
\label{masterGA}
\frac{dN_k}{dN}=&\frac{(1\!-\!f_{k\!-\!1})N_{k-1}-(1\!-\!f_k)N_k}{N} 
+\frac{(k\!-\!2)t_{k-1}N_{k\!-\!1}-(k-1)t_{k}N_{k}}{N} \\
&~~~~~~~~~~~~+\frac{g_{k\!-\!1}u_{k-1}N_{k\!-\!1}-g_{k}u_{k}N_{k}}{N}+\delta_{k,1}\,.
\end{split}
\end{align}
The first two ratios are the same as in Eq.~\eqref{masterRA}, and correspond
to situations where an incoming node links to the target or to the parent of
the target.  The third ratio is specific to grandparent redirection and
corresponds to the situation where a link is redirected to the grandparent of
a target node.  For example, $g_ku_kN_k/N$ is the probability that one of the
$g_kN_k$ grandchildren of degree $k$ nodes is initially targeted and the link
is redirected to the degree-$k$ grandparent.

\begin{figure}[ht]
\begin{center}
\includegraphics[width=0.55\textwidth]{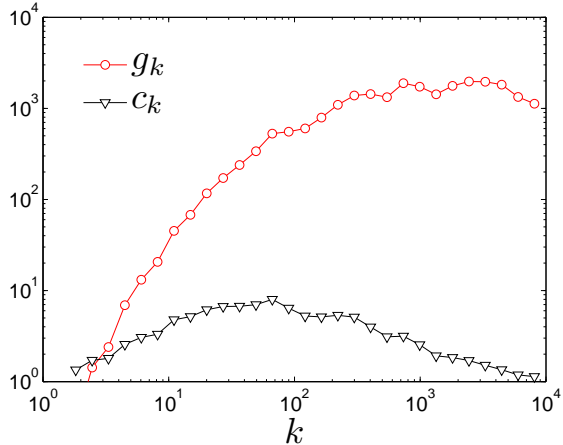}
\caption{Dependence of $g_k$ ($\circ$) and $c_k$ ($\bigtriangledown$) versus
  $k$ from direction simulations of superlinear preferential attachment
  growth for $\gamma=1.3$.  Data represents $10^3$ network realizations of
  size $N=10^4$.  Qualitatively similar results are obtained for other values
  of $\gamma>1$. }
\label{dk}
\end{center}
\end{figure}

Rearranging terms and eliminating $g_k$ in favor of $c_k$, we may express the
master equation in the canonical form of~\eqref{masterPA}, with the
attachment rate
\begin{equation}
\frac{A_k}{A}=\frac{(k-1)(c_k-1)u_k+(k-1)t_k+1-f_k}{N}~.
\label{GR}
\end{equation}
If $c_k$ grew as a power law in $k$, then the network would grow according to
superlinear preferential attachment.  However, we can show that $c_k$ must be
bounded as $k$ increases by considering the hubs of the network.  A key
feature of superlinear preferential attachment networks, is the presence of a
single hub node whose degree is of the order of
$N$~\cite{GBW04,OS05,KRB10,KK08}.  From the relation $g_k=(k-1)(c_k-1)$, we
see that the hub has $g_N\sim c_N\times N$ grandchildren.  If $c_k$ grows
without bound, then for large network size $N$, the hub will have more
grandchildren than total number of nodes in the network, which is impossible.
Therefore, $c_k$ must be be bounded for large $k$ and grandparent redirection
cannot produce superlinear preferential attachment.

Indeed, direct simulations of superlinear preferential attachment networks of
up to $N=10^4$ nodes show that $c_k$ asymptotically approaches 1 as $k\to N$
(Fig.~\ref{dk}).  Typically, the hub node connects to order $N$ `leaf' nodes
of degree $1$.  Moreover, the number of nodes with degrees larger than, but
of the order of, one grows at most sublinearly with $N$~\cite{KR01}.
Consequently, the mean degree of a child of the hub asymptotes to $1$.
Similarly, as their degree increases, non-hub nodes will be connected to a
larger fraction of leaves.  Thus $c_k$ decreases with $k$ for large $k$, so
that the average number of grandchildren, $g_k=(k-1)(c_k-1)$, grows
sublinearly with $k$ (Fig.~\ref{dk}).

Grandparent redirection illustrates a basic shortcoming of any local growth
rule in the quest to produce superlinear preferential attachment network
growth.  Consider a arbitrary node $\mathbf{x}$ in a network that grows by
some local redirection attachment rule.  The rate at which a new node
attaches to $\mathbf{x}$ is limited by the size of its local environment, now
defined as the set of nodes that, if initially targeted by an incoming node,
can ultimately lead to attachment to $\mathbf{x}$.  For nearest-neighbor
redirection, the local environment of $\mathbf{x}$ consists of $\mathbf{x}$
and its children, and has size $k$ if $\mathbf{x}$ has degree $k$.  In
grandparent redirection, the local environment of $\mathbf{x}$ consists of
its children and grandchildren, so that its size equals $k+g_k$.  Because
$g_k$ grows sublinearly in $k$, grandparent redirection cannot give
superlinear preferential attachment.

To achieve superlinear preferential attachment, the size of the local
environment of a node must grow superlinearly in the node degree.  For a hub,
this condition leads to the impossible situation that the local environment
must have a size of order $N^\gamma$, which is larger than the entire
network.  


\section{Conclusion}
We presented a degree-dependent nearest-neighbor redirection algorithm to
generate complex networks.  We showed how this algorithm is equivalent to
network growth by sublinear preferential attachment when the redirection
probability is a suitably-chosen function of the degrees of the target and
parent nodes.  By exploiting only this local information in the immediate
vicinity of the target node, we constructed sublinear preferential attachment
networks extremely efficiently because just a few computer instructions are
needed to create each new network node.  We also argued that no local
redirection rule can generate superlinear preferential attachment.  The
prevalence of linear and sublinear preferential attachment networks, along
with the relative scarcity of superlinear preferential attachment networks,
suggests that real-world networks should grow only according to local growth
rules.

\ack We gratefully acknowledge financial support from grant
\#FA9550-12-1-0391 from the U.S. Air Force Office of Scientific Research
(AFOSR) and the Defense Advanced Research Projects Agency (DARPA).

\end{document}